\documentclass[graybox]{svmult}

\usepackage{mathptmx}       
\usepackage{helvet}         
\usepackage{courier}        
\usepackage{type1cm}        
\usepackage{makeidx}         
\usepackage{graphicx}        
\usepackage{multicol}        
\usepackage[bottom]{footmisc}

\makeindex             

\newcommand{\simgreat} {\mathbin{\lower 3pt\hbox{$\rlap{\raise
        5pt\hbox{$\char'076$}}\mathchar"7218$}}}

\newcommand{\simless}{\mathbin{\lower 3pt\hbox {$\rlap{\raise
        5pt\hbox{$\char'074$}}\mathchar"7218$}}}

\begin{document}

\title*{Spectropolarimetry of stars\\across the H-R diagram}
\author{Swetlana Hubrig \and Markus Sch\"oller}

\institute{
Swetlana Hubrig \at Leibniz-Institut f\"ur Astrophysik Potsdam (AIP), An der Sternwarte~16, 
14482~Potsdam, Germany, email: {shubrig@aip.de}\\
Markus Sch\"oller \at European Southern Observatory, Karl-Schwarzschild-Str.~2, 85748~Garching, Germany,
email: {mschoell@eso.org}}
%
%
\maketitle

\abstract{
The growing sample of magnetic stars shows a remarkable diversity
in the properties of their magnetic fields. The overall goal of current studies is to 
understand the origin, evolution, 
and structure of stellar magnetic fields in stars of different mass at different evolutionary stages. 
In this chapter we discuss recent measurements together with the underlying assumptions in 
the interpretation of data and the requirements, both observational and theoretical, for obtaining 
a realistic overview of the role of magnetic fields in various types of stars.}

\section{Introduction}
\label{sec:1}

The presence of a magnetic field in the Sun was detected in 1908 by Hale
and the first detection of a magnetic field in a distant star was achieved in 1947 by Babcock. 
Since then, a lot of effort has been put into investigations of solar and stellar magnetism to 
understand how and why so many stars, including our own Sun, are magnetic, and what the 
implications might be for life on Earth and elsewhere in the universe.
The past two decades have seen a significant step forward in our understanding of the occurrence of magnetic fields in 
stars of different mass and different age. This became possible through the significant progress in 
spectropolarimetric surveys of surface magnetic fields that made use of the currently available 
spectropolarimeters 
ESPaDOnS (Echelle SpectroPolarimetric Device for the Observation of Stars) at the Canada-France-Hawaii 
telescope, NARVAL at the Bernard Lyot 2\,m Cassegrain telescope, the Main Stellar Spectrograph of the 
Russian 6\,m telescope,
FORS\,1 and FORS\,2 (FOcal Reducer low dispersion Spectrographs) at the ESO Very Large Telescope (VLT), and 
HARPS\-pol (High Accuracy Radial velocity Planet Searcher) at ESO/La~Silla.  

As of today, magnetic fields are found at all stages of stellar evolution, 
from brown dwarfs and young T\,Tauri stars to the end products of stellar evolution: white dwarfs and neutron 
stars. The incidence of magnetic fields in stars is diverse. The presence of a convective envelope is a 
necessary condition for significant magnetic activity in stars and magnetic activity is found all the way 
from late A-type stars, e.g.\ in Altair (Robrade \& Schmitt 2009) with very 
shallow convective envelopes, down to the 
coolest fully convective M-type stars. All solar type stars appear to be magnetic, the stronger the more 
rapidly they rotate (e.g.\ Pallavicini et al.\ 1981).
This is understood through the $\alpha\Omega$-dynamo, which is thought to 
operate in the convective envelope of these stars (e.g.\ Schmitt 1987). In intermediate mass main 
sequence stars, only 
about 10\% are found to have kG-strength large-scale magnetic fields. Here, the correlation of magnetic 
fields with stellar rotation is opposite to that of solar type stars: whereas most intermediate mass 
stars are rapid rotators throughout their main sequence life, the magnetic stars are mostly slow rotators
(e.g., Mathys 2017; Hubrig et al.\ 2018a). 

In spite of the fact that recent years have seen a significant increase in observational studies of 
stellar magnetism, 
the most important aspects, such as the origin of stellar magnetic fields and the evolution of magnetic field 
configurations -- the large-scale organised magnetic fields observed in massive stars or the complex magnetic 
topology observed in low-mass stars -- are still not understood. 

For massive and intermediate mass stars with radiative envelopes, it has been argued
that their magnetic fields could be fossil relics of the fields that were present in the interstellar medium 
from which these stars have formed (e.g.\ Moss 2003). A search for the presence of magnetic fields
in massive stars located in active sites of star formation, in the $\rho$\,Ophiuchus star-forming cloud and
in the Trifid nebula, led to the detection of magnetic fields of several kG in two early 
B-type stars, the B2\,V star HD\,147933 and the B1\,V star HD\,164492Cb (Hubrig et al.\ 2018b, 2014a).   
However, the fossil field hypothesis has several problems as it does not explain the low ($\sim10$\%) 
occurrence of magnetic stars and their broad range of field strengths. 
Moreover, a study of the distribution of magnetic Ap stars in the H-R diagram using accurate Hipparcos 
parallaxes by Hubrig et al.\ (2000) 
showed that these stars are concentrated toward the centre of the 
main-sequence band and only rarely can be found close to the zero-age main sequence (ZAMS). 
Similar studies using Gaia DR2 data are on the way, but their results are not published yet.
Alternatively, magnetic fields may be 
generated by strong binary interaction, i.e., in stellar mergers, or during a mass transfer or common 
envelope evolution (e.g.\ Tout et al.\ 2008). The resulting 
strong differential rotation is considered as a key ingredient 
for the generation of magnetic fields (Petrovic et al.\ 2005). For Galactic 
O-type stars, it was shown that the fraction of very close 
binaries is so high that about one third of them interacts with a companion during their main sequence 
evolution through mass transfer or merging (Sana et al.\ 2012). 

In cool, low-mass stars, the key unresolved aspects are the origin of 
the solar-like activity cycle, the origin and character of differential rotation and its role in the large-scale 
dynamo, and the origin of high energy coronal activity.
Knowledge of the evolution of the magnetic field and its configuration is extremely important for our 
understanding of the past and future of the Sun and the solar system. Notably, low-mass stars are complex 
magnetohydrodynamical objects that require state-of-the-art observations and numerical simulations 
to pin down the physical processes that are responsible for the diversity and dynamics of these stars.
The origin of stellar activity in low-mass stars resides most likely in the stellar convection zones. 
The mechanism involves helical gas motions and large-scale differential rotation. The most straightforward 
combination of these effects is the $\alpha\Omega$-dynamo, which can reproduce many basic properties of the solar 
activity cycle. Differential rotation generates large-scale toroidal fields by “winding up” the poloidal 
field. Differential rotation is the main field generator in the $\alpha\Omega$-dynamo, which therefore maintains predominantly 
axisymmetric fields that are dominated by the toroidal component. On the other hand, differential rotation 
alone is not sufficient for dynamo action, as it can not prevent the decay of the poloidal field. 
The $\alpha$-effect, which is caused by the Coriolis force, is vital to the dynamo, because it produces 
a poloidal field from the toroidal field, thus completing the field-amplifying dynamo process. In contrast 
to differential rotation, the $\alpha$-effect can maintain a magnetic field by itself. This type of dynamo 
is known as the $\alpha^{2}$-dynamo. Unlike the $\alpha\Omega$-dynamo, it prefers non-axisymmetric field 
configurations that are not dominated by one component. Mean field magnetohydrodynamics provides a 
very successful theoretical model for stellar differential rotation and the accompanying meridional 
flows (Kitchatinov \& R\"udiger 1999; K\"uker \& Stix 2001). 

In the following sections,
we discuss magnetic field diagnostics,
observations of magnetic fields in upper main-sequence,
pre-main-sequence, and low-mass active stars,
finishing with post-main-sequence stars and degenerate stellar remnants.

\section{Magnetic field diagnostics}
\label{sec:2}

The currently applied observing techniques differ for different types of stars, involving the Zeeman effect, 
the Paschen-Back effect, and the molecular Zeeman effect. Observations of broad-band linear polarization,
which is caused by different saturation levels of the $\pi$ and $\sigma$ components of
a spectral line in the presence of a magnetic field, were reported in the past for a few chemically peculiar magnetic 
Ap stars (e.g.\ Leroy 1989; Landolfi et al.\ 1993). This differential
effect  for Zeeman components is qualitatively similar for all lines, so that in broad-band
observations the contributions of all lines add up. Monitoring of broad-band linear polarization of 
a few Ap stars over a few years was used to establish the presence 
of their extremely slow rotation. Studies of broad-band linear polarization do not represent a widely 
used technique of stellar magnetic field diagnosis, and nowadays they are superseded by the observations 
of linear polarization in individual spectral lines.

The most complete way to study stellar magnetic fields is to observe both linear and circular 
polarization, i.e.\ all four Stokes parameters $I$, $Q$, $U$, and $V$. However, such observations 
are very challenging as 
in disk-integrated observations the signal can be canceled or significantly decreased due to mixed polarity 
fields. Furthermore, measurements of linear polarization are sensitive to instrumental polarization.
Therefore, only large-scale magnetic fields can be studied using the full Stokes vector. Importantly, 
physically realistic models of the geometric structure of magnetic fields in stellar atmospheres 
can be achieved only if linear polarisation measurements are implemented.

Observing the spectrum of a star in the two senses of circular polarisation
is in almost all cases the most sensitive method of detecting weak magnetic fields on a stellar surface. 
The measurement of the Stokes~$V$ parameter permits to detect the mean longitudinal magnetic field,
which is a line intensity weighted average over the visible hemisphere of a star of the magnetic field component 
directed along the line-of-sight. To measure longitudinal magnetic fields, in particular observing with 
low-resolution spectropolarimeters, a regression analysis approach is frequently used (e.g.\ Bagnulo et al.\ 2001).
The mean longitudinal magnetic field measurement $\left< B_{\rm z}\right>$ in this approach is  
based on the relation

\begin{eqnarray} 
\frac{V}{I} = -\frac{g_{\rm eff}\, e \,\lambda^2}{4\pi\,m_{\rm e}\,c^2}\,
\frac{1}{I}\,\frac{{\rm d}I}{{\rm d}\lambda} \left<B_{\rm z}\right>\, ,
\label{eqn:vi}
\end{eqnarray} 

\noindent 
where $V$ is the Stokes parameter that measures the circular polarization, $I$
is the intensity in the unpolarized spectrum, $g_{\rm eff}$ is the effective
Land\'e factor, $e$ is the electron charge, $\lambda$ is the wavelength,
$m_{\rm e}$ is the electron mass, $c$ is the speed of light, 
${{\rm d}I/{\rm d}\lambda}$ is the wavelength derivative of Stokes~$I$, and 
$\left<B_{\rm z}\right>$ is the mean longitudinal magnetic field.
Although the longitudinal magnetic field 
measurements present the canonical diagnostics of stellar magnetic fields, other observational and 
analytical methods to diagnose the presence and structure of magnetic fields in stellar atmospheres are
frequently applied. 

In a number of Ap and Bp stars whose projected rotational velocity is sufficiently small and 
whose magnetic field is strong enough to exceed the rotational Doppler broadening, certain lines 
with suitable Zeeman patterns show resolved Zeeman split lines in unpolarized light 
(Preston 1971; Mathys 1990). They are 
used to estimate the line-intensity weighted average of the modulus of the magnetic field over the visible
stellar hemisphere, the mean magnetic field modulus. Simultaneous knowledge of the 
mean magnetic field modulus and the longitudinal magnetic field over the stellar rotation period permits
to set constraints on the magnetic field geometry.

Since the Zeeman signatures in the spectra of many stars are generally very small, and increasing the 
signal-to-noise ratio ($S/N$) by increasing the exposure time is frequently limited by the short 
rotation period of the star, multi-line 
approaches as proposed by Semel (1989) are commonly used to increase the $S/N$. The most widely used 
approach is the Least Squares Deconvolution (LSD; Donati et al.\ 1997). The main assumption in LSD is the
application of the weak field approximation, that is, the magnetic splitting of spectral lines is 
assumed to be smaller than their Doppler broadening. Furthermore, it is assumed that the local line profiles 
are self-similar and can be combined into an average profile. Due to non-linear effects in the summation 
and the effect of blends, the resulting LSD profiles should not be considered as observed, but rather 
processed Zeeman signatures.
In a number of recent studies, a novel magnetometric technique, the multi-line Singular Value 
Decomposition (SVD) method was applied.
The software package to study weak magnetic fields using 
the SVD method was introduced by Carroll et al.\ (2012).
The basic idea of SVD is similar to the principal component analysis (PCA) approach, where the 
similarity of individual Stokes~$V$ profiles permits to describe the most coherent and 
systematic features present in all spectral line profiles as a projection onto a small number 
of eigenprofiles. 

Other popular methods to exploit the information contained in polarised spectral line profiles 
currently include the moment technique, in particular the determination of the crossover effect and of the 
mean quadratic field (e.g.\ Mathys 1993), and Zeeman-Doppler imaging (ZDI). In the moment technique, the shapes
of the spectral lines observed in Stokes~$I$ and Stokes~$V$ profiles can be characterized by their moments
of various orders about the line centre and can be interpreted in terms of quantities related to the 
magnetic field. The analysis is usually based on the consideration of samples of reasonably 
unblended lines and critically depends on the number of lines that can be employed. 
ZDI is an inversion technique applied to time-series of two or four Stokes parameters,
introduced by Semel (1989). It allows to recover the distribution of the temperature (on the surface of cool stars) 
or of chemical spots (on the surface of Ap and Bp stars) and of the magnetic field vector over the stellar surface.
In most studies, however, only Stokes~$I$ and Stokes~$V$ observations are used to obtain 
Zeeman-Doppler stellar images.
Obviously, in such cases, the solution is not unique and is constrained by a number of assumptions.
In the studies of cool stars with temperature spots it is however possible to remedy this situation
by the implementation of full polarized radiative transfer in both atomic and molecular lines.   Employing
molecular lines is essential to localize magnetic fields inside cool regions on the stellar surface 
(e.g.\ Sennhauser et al.\ 2009).

A novel technique to detect surface magnetic fields in A- and F-type stars, which uses the autocorrelation 
of spectra in integral light, was recently suggested by Borra \& Deschatelets (2015). 
The original idea of the autocorrelation technique is to measure the width of spectral lines and to disentangle the 
magnetic broadening from other broadening effects, using the differences in wavelength dependence. 
Such a technique can be applied to low-resolution spectra, down to stars that are too faint for 
high-resolution observations even with large telescopes. The availability of this technique allows one
to  use publicly available data archives from numerous spectrographs all over the world to obtain 
reliable statistics on the occurrence of magnetic fields in A and F stars in clusters and the field 
at different evolutionary stages. Using this technique, it will become possible even to search for magnetic 
fields in a sample of early B-type stars in the Magellanic Clouds using low-resolution spectroscopic
observations.

\section{O- and early B-type stars}
\label{sec:3}

Massive stars with initial masses ranging from 8 to 100\,$M_{\odot}$ are of particular interest, 
as they end their evolution with a supernova explosion, producing neutron stars or black holes. 
Magnetic O-type stars with masses larger than 30\,$M_{\odot}$ and their WR descendants have been 
suggested as progenitors of magnetars (Gaensler et al.\ 2005). A magnetic mechanism for the collimated explosion of 
massive stars, relevant for long-duration gamma-ray bursts, X-ray flashes, and 
asymmetric core collapse supernovae was proposed by e.g.\ Uzdensky \& MacFadyen (2006).

Recent observations indicate that probably only a small fraction of about 7$\pm$3\% of O-type 
stars with masses exceeding 18~$M_{\odot}$ (Grunhut et al.\ 2017)
and about 6$\pm$3\% of early B-type and O-type stars (Sch\"oller et al.\ 2017)
possess measurable, mostly dipolar magnetic fields.
Progress in the studies of massive stars became possible mostly through the 
MiMeS (The Magnetism in Massive Stars, Wade et al.\ 2011), MAGORI (MAGnetic field ORIgin, Hubrig et al.\ 2011), 
and BOB (B-fields in OB stars, Morel et al.\ 2015) surveys, which
made use of the spectropolarimeters ESPaDOnS, NARVAL, FORS2, and HARPS\-pol. 
The vast majority of magnetic massive stars exhibit a smooth, single-wave variation
of the longitudinal magnetic field during the stellar rotation cycle.  These  observations are considered  
as  evidence for a dominant dipolar contribution to the magnetic field topology.

Braithwaite \& Spruit (2004) showed that stars with 
radiative envelopes can host stable large-scale magnetic 
field configurations, as those observed in the Ap/Bp stars, independent from the star's rotation. The implication 
is that up to about 30-50\% of all massive stars might be magnetic, and that so far only the tip of the iceberg 
has been seen (Stello et al.\ 2016).
From Kepler and Kepler K2 photometry, Balona (2013, 2016) 
concluded that about 40\% of A-type stars and 30-50\% of late O and B-type stars show rotational 
modulation due to the presence of star spots. Pulsating stars, eclipsing binaries, and ellipsoidal 
variables of high amplitude were avoided in these studies. Since it is difficult to conceive of 
starspot formation without the presence of a magnetic field, these results indicate that the 
occurrence of magnetic fields in A, B, and late O-type stars is greatly underestimated. 

No complete volume-limited study of the incidence of magnetism in massive stars has been presented so far.
Thus one of the major goals of recent studies 
is to build trustworthy statistics on the occurrence of magnetic fields in 
massive stars and their topology at different evolutionary stages. This is critical to answer the 
principal question of the possible origin of such fields. Previously observed samples greatly suffer 
from a number of biases in the target selection related to specific spectral classifications. We know already 
that all five Galactic stars with Of?p classification are magnetic and at least 10\% of the early-type 
B stars belong to the magnetic He-rich group, containing stars with overabundances of Si and He. 
O-type stars with Of?p classification 
exhibit recurrent, and apparently periodic, spectral variations in Balmer, He~{\sc i}, C~{\sc iii}, and Si~{\sc iii}
lines, sharp emission or P~Cygni profiles in He~{\sc i} and the Balmer lines, and strong C~{\sc iii} emission
around 4650\,\AA{} (Walborn 1972).
A search for the presence of magnetic fields at 
different evolutionary stages is also predominantly carried out in bright stars that are slow rotators. 
Since very young massive stars at the ZAMS are usually faint, this evolutionary 
stage of massive stars remains completely unexplored. The results recently presented in several 
papers reflect the lack of observations of young massive stars close to the ZAMS (e.g.\ Castro et al.\ 2015). 

It is quite possible that a significant fraction of the magnetic massive stars remains undiscovered due 
to a more complex magnetic field topology related to the presence of a subsurface convection zone. 
Multi-wavelength observations of massive stars show cyclical line profile structures, 
the discrete absorption components (DACs). Their presence indicates that stellar prominences can emerge on the 
stellar surface, probably due to localised magnetic activity, more or less in analogy to what is 
observed on the solar surface (e.g.\ Sudnik \& Henrichs 2016). 
Action of a subsurface convection zone (Cantiello et al.\ 2009) 
would be the most likely driving mechanism that generates magnetic spots, which are the source of those prominences. 
This zone may also be the source of a global magnetic field, winding up toroidally with stochastic 
buoyancy breakouts, causing corotating magnetic bright spots at the surface of 
the star (Cantiello \& Braithwaite 2011).

Atmospheric magnetic fields can impact surface rotation rates via magnetic braking
(Weber \& Davis 1967; ud-Doula et al.\ 2008), introduce chemical abundance inhomogeneities and 
peculiarities (Hunger \& Groote 1999), and confine the stellar wind in a magnetosphere 
(e.g., Friend \& MacGregor 1984; ud-Doula \& Owocki 2002). 
The ejected matter streams along the field lines towards the magnetic equator and 
collides in the vicinity of the equatorial
plane, leading to strong shocks and thereby heating plasma to
X-ray temperatures (Babel \& Montmerle 1997).

 The strongest longitudinal magnetic field in an O star of more than  5\,kG was detected in the Of?p star NGC\,1624-2,
suggesting a dipole component of about 20\,kG (Wade et al.\ 2012). Among the early B-type stars, the record holder
is the He-rich star CPD\,$-$62$^{\circ}$\,2124 with a dipole component of at least 21\,kG (Hubrig et al.\ 2017).

Noteworthy, the survey of the BOB consortium, aiming to establish a lower boundary of 
magnetic fields in massive stars, revealed the presence of very weak Zeeman signatures across 
the line profiles in high-quality HARPS\-pol spectra of two early B-type stars. This work 
indicated that also magnetic 
fields of the order of a few Gauss can be present in massive stars, but possibly escape their detection due to 
insufficient measurement accuracy (e.g.\ Fossati et al.\ 2015). 

Wolf-Rayet (WR) stars form an integral part of the late evolution of the most massive stars. 
While the WR core He-burning phase lasts only $\sim10$\% of the O-star main sequence phase,
WR winds are at least ten times 
stronger than on the main sequence. With wind-momenta in WR stars reaching up to 30 times the available 
photon momentum, magnetic fields have been proposed as additional or alternative wind-driving mechanisms 
to radiation pressure (Cassinelli et al.\ 1995). The detection of magnetic fields in WR stars 
is however difficult, chiefly because the line spectrum is formed in the strong stellar wind. This does 
not only imply a dilution of the field at the place of line formation. The biggest problem is the wind 
broadening of the emission lines by Doppler shifts with wind velocities of a few thousand km\,s$^{-1}$. 
de la Chevroti\'ere et al.\ (2014) used  ESPaDOnS spectra to search for magnetic fields in eleven WR stars.
They reported that the data show evidence supporting marginal detections of a magnetic field of the
order of a few hundred Gauss and less in three WR stars. A search 
for magnetic fields in several WR stars using low-resolution FORS\,2 spectropolarimetry 
showed the presence of a  magnetic field in the cyclically variable star WR\,6 at a significance level of 
3.3$\sigma$ (258$\pm$78\,G). In this work, a theoretical approach to estimate 
the magnetic field strength in stars with strong emission lines was considered for the first time 
(Hubrig et al.\ 2016).

\section{Ap and Bp stars}
\label{sec:4}

Globally ordered magnetic fields are observed in roughly 10\% of the
intermediate and massive main-sequence stars with spectral types between 
approximately B2 and F0. These stars, generally called chemically peculiar 
Ap and Bp stars, exhibit strong overabundances of certain elements,
such as iron peak elements and rare earths,
and underabundances of He, C, and O, relative to solar 
abundances, and are characterized observationally 
by anomalous line strengths.  Massive Bp stars usually
show overabundances of He and Si. As the star rotates, the magnetic field
and surface abundance distribution is observed from various aspects,
resulting in variability of the measured magnetic field and spectral line strengths.

The variable magnetic fields are generally diagnosed through
mean longitudinal magnetic field, mean magnetic field modulus, and net
broadband linear polarization measurements. Chemical abundance anomalies are commonly believed to be
due to radiatively driven microscopic diffusion in stars rotating
sufficiently slowly to allow such a process to be effective (e.g.\ Michaud 1970). 
While two thirds of all Galactic O-type stars appear to be members of close binary systems 
(Sana et al.\ 2012), the study of Carrier et al.\ (2002) indicated 
a dearth of magnetic Ap stars stars 
in close binaries. Only for two close binary systems with Ap magnetic components, HD\,98088 
(e.g., Babcock 1958; Abt et al.\ 1968) and HD\,161701 (Hubrig et al.\ 2014b), is the structure of 
the magnetic field known. Remarkably, in both systems carries the 
surface of the Ap component, which is facing the companion, a positive magnetic field. It is, however, 
not clear whether tidal forces may play a role during the dynamic process of tidal 
synchronisation, to align the field geometries in the observed way. Further systematic searches 
for magnetic fields in binary systems should be conducted to properly evaluate theoretical models 
of the origin of their magnetic fields.

\begin{figure}[t]
\centering
  \includegraphics[width=\textwidth]{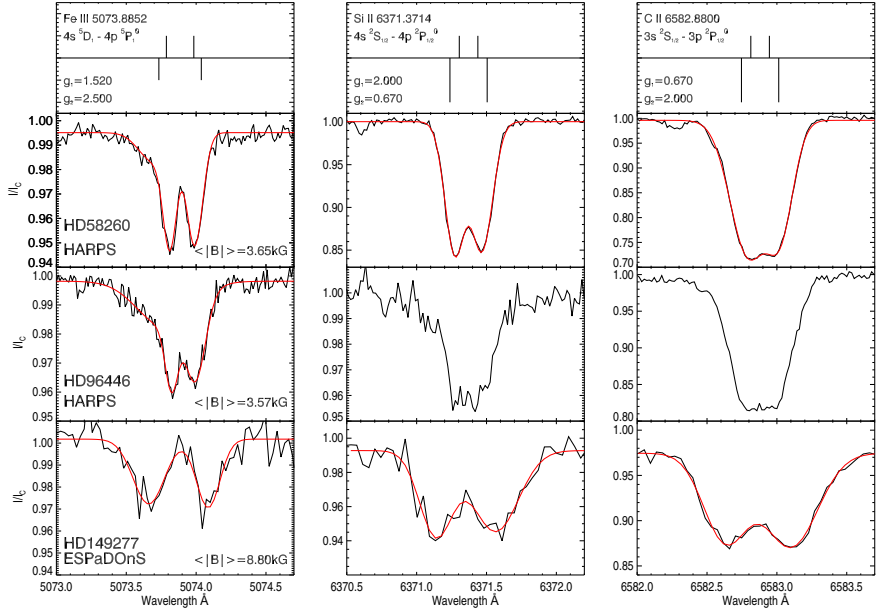}
\caption{The magnetically split lines Fe\,{\sc iii}~$\lambda$5074, Si\,{\sc ii}~$\lambda$6371, and 
C\,{\sc ii}~$\lambda$6583 in 
high-resolution Stokes~$I$ spectra of the early B-type stars HD\,58260, HD\,96446, and HD\,149277.
The red lines denote the fit of a multi-Gaussian to the data.
For two lines in the spectrum of HD\,96446, the splitting is not sufficient to allow a proper fit.
}
\label{fig:res}
\end{figure}

Currently, 84 stars with magnetically
split components are known and their study permits to establish
the general properties of their magnetic fields (e.g.\ Mathys 2017). The star 
with the strongest magnetic field currently known, ``Babcock's star'' 
HD\,215441 is a B8\,V star with a surface dipole magnetic field strength of 34\,kG 
(Babcock 1960; Preston 1969). 
Zeeman split lines were recently discovered in a few early Bp stars with low projected
rotational velocity and strong, kG order magnetic fields (Hubrig et al., in prep.).
Examples of magnetically split lines in three such stars are presented in Fig.~\ref{fig:res}. 

The evolution of magnetic field configurations in stars with masses  
below 6~$M_{\odot}$ using a sample of Ap and Bp stars with accurate Hipparcos parallaxes and 
definitely determined longitudinal magnetic fields, was until 
now considered only in the study by Hubrig et al.\ (2007). This work showed that stronger magnetic fields 
are detected in younger stars. No evidence was detected for any loss 
of angular momentum during the main-sequence life. Also the magnetic flux was found to be constant 
over the stellar life time on the main sequence. An excess of stars with large obliquities 
$\beta$ (the orientation of the magnetic axis with respect
to the rotation axis) was detected in both higher and lower mass stars. 
The evolution of $\beta$ appeared 
to be somewhat different for stars with M$<$3\,$M_{\odot}$ than for stars with 
M$>$3\,$M_{\odot}$, for which a strong hint for an increase of $\beta$ with time spent on the 
main sequence was discovered. 

Using the indirect surface mapping ZDI method is a valuable option to derive magnetic field maps, but was applied
only for a few Ap/Bp stars during the last years. A summary of ZDI results is presented in the work of 
Kochukhov et al.\ (2018). The authors find that the global magnetic field geometry changes little from
one star to another, with nearly all stars showing dominant dipolar fields with a varying degree
of distortion.

\section{Pre-main sequence stars}
\label{sec:5}

Studies of magnetic fields in stars at early evolutionary stages, before they arrive on the main sequence, 
are of special interest to get an insight into the magnetic field origin. 
Moreover, such studies are extremely important because they enable us to 
improve our insight into how the magnetic fields in these stars are generated and how they interact with 
their environment, including their impact on the planet formation process and the planet-disk interaction.

It is generally accepted that 
magnetic fields are important ingredients of the star formation process 
(e.g.\ McKee \& Ostriker 2007) and are already 
present in stars in the pre-main sequence (PMS) phase.  The PMS low mass 
T\,Tauri stars stand out by their very large 
emissions in chromospheric and transition-region lines. The accretion of circumstellar disk material 
onto the surface of these stars is controlled by a strong stellar magnetic field. 
These stars are
usually grouped into two classifications: classical T\,Tauri stars, which show evidence of a
circumstellar disk and mass accretion onto the central star in the form of excess emission in X-rays,
the UV, optical, and infrared, and weak-line T\,Tauri stars, which 
do not show evidence for significant mass accretion.  
A model for the generation of magnetic fields in 
fully convective PMS stars was presented by e.g., K\"uker \& R\"udiger (1999),
Chabrier \& K\"uker (2006), and  K\"uker (2009), who considered  a 
dynamo process of $\alpha^{2}$-type. 

The accretion of circumstellar disk material 
onto the surface of T\,Tauri stars is assumed to be controlled by a strong stellar magnetic field. 
Furthermore, the magnetic field appears to be critical for explaining their rotational properties 
(e.g.\ Herbst et al.\ 2007) and probably plays a critical role in driving the observed outflows 
(Mohanty \& Shu 2008). Magnetic field measurements of these stars over the last decades show the presence
of kG magnetic fields. Basri et al.\ (1992) were the first to detect a magnetic field on the surface
of a T\,Tauri star using Zeeman broadening of magnetically sensitive lines in Stokes~$I$.
Modeling Zeeman-broadened Stokes~$I$ profiles often involves template spectra of magnetically active and
inactive stars, which are weighted by a filling factor.
As this method is sensitive to the magnetic field distribution but has a limited sensitivity to the magnetic 
field geometry, most recent studies involve 
circular polarization observations to
characterize surface magnetic fields using the ZDI technique. On the other hand, according to
Carroll et al.\ (2012),  a common problem for ZDI is that the real strong magnetic
fields that are associated with cool surface regions produce only a fraction of the photon flux 
compared to the unaffected quiet or even hot surface regions. 
When  studying the correlation between  the magnetic field
and temperature spots, it is important to be aware that the detectability
of the field is weighted by the surface brightness. 
To  prevent such an
ill-defined flux-weighting and the mutual effects between
the temperature and magnetic field, Carroll et al.\ have pursued a strategy
that performs a simultaneous DI and ZDI to retrieve both the
temperature and the magnetic field distribution at the same time.
Their reconstruction of the temperature and magnetic field of the weak-line T\,Tauri star
V410\,Tau revealed that the majority of the strong fields of 2\,kG
are located in cool spots, in particular within the large polar spot. The reconstruction 
clearly showed that both polarities coexist within the large polar spot and that the entire polar-field
topology appears to be dominated by a twisted bi-polar structure separated by a relatively 
sharp neutral line. 

To study magnetic field topologies in T\,Tauri stars, the group of Donati
(e.g.\ Donati et al.\ 2015) applies a stellar-surface tomographic-imaging package.
This imaging code is set up to invert (both automatically
and simultaneously) time series of Stokes~$I$ and $V$ LSD profiles
into brightness and magnetic maps of the stellar surface, where
brightness imaging is allowed to reconstruct both cool spots
and warm plages, known to contribute to the activity of very active stars. 
This tool is
based on the principles of maximum-entropy image reconstruction
and on the assumption that the observed variability is mainly caused
by rotational modulation with an added option for differential rotation.

According to Adams \& Gregory (2012), in accreting systems, the polar strength of the dipole component
is particularly relevant for determining the disk truncation radius and the balance of torques 
in the star-disk system. Johnstone et al.\ (2014) found that the dipole strength is correlated
with the rotation period: stars with weak dipole components tend to be rotating
faster than stars with strong dipole components. 
Recent observations of magnetic fields in a small sample of T\,Tauri stars suggest that the magnetic field topology 
changes as a function of age, i.e.\ there is a relation
between the large-scale magnetic field and the position in the H-R diagram.
Gregory et al.\ (2012) reported that the  magnetic field  gains  complexity with the evolution of the star:
the magnetic field is mainly dipolar when the star is completely convective. After the
development of the radiative core the dipolar component looses power compared to high
order components of the multipole expansion.
Based on magnetohydrodynamic simulations, Zaire et al.\ (2017) suggest that the presence 
of a tachocline could be an important factor in the development of magnetic fields with higher 
multipolar modes. 

The presence of magnetic fields in higher mass PMS stars, the Herbig Ae/Be stars, 
has long been suspected, in particular on account of H$\alpha$ spectropolarimetric observations pointing out the 
possibility of magnetospheric accretion, similar to that of classical T\,Tauri stars (e.g.\ Vink et al.\ 2002).
While models of magnetically driven accretion and outflows successfully reproduce many observational 
properties of the classical T\,Tauri stars, the picture is completely unclear for 
the Herbig Ae/Be stars, mostly due to the poor knowledge of their magnetic field topology. 
So far, the magnetic field geometry was constrained only for two Herbig Ae/Be stars, 
V380\,Ori (Alecian et al.\ 2009) and HD\,101412 (Hubrig et al.\ 2011), 
and only about 20 Herbig stars were reported to host magnetic fields 
(Hubrig et al.\ 2015, and references therein). 

The two best studied Herbig Ae/Be stars exhibit a single-wave variation in their mean longitudinal 
magnetic field during the stellar rotation cycle. This behaviour is usually considered as evidence 
for a dominant dipolar contribution to the magnetic field topology.
Presently, the Herbig Ae star HD\,101412 possesses the strongest 
magnetic field ever measured in any Herbig Ae star, 
with a surface magnetic field $\left<B\right>$ of up to 3.5\,kG. 
HD\,101412 is also the only Herbig Ae/Be star for which the rotational
Doppler effect was found to be small in comparison to the magnetic splitting, presenting
several spectral lines resolved into magnetically split components
observed in unpolarised light at high spectral resolution (Hubrig et al.\ 2010).

Notably, the task of magnetic field measurements in Herbig stars is very challenging, 
as the work of Hubrig et al.\ (2015) demonstrates, in which the authors 
compiled all magnetic field measurements reported in 
previous spectropolarimetric studies. This study indicates that the low detection rate of magnetic 
fields in Herbig Ae stars, about 7\% (Alecian et al.\ 2013), can indeed be 
explained not only by the limited sensitivity of 
the published measurements, but also by the weakness of these fields. The obtained density distribution 
of the rms longitudinal magnetic field values reveals that only a few stars have magnetic fields stronger than 
200\,G, and half of the sample possesses magnetic fields of about 100\,G and less. 
Consequently, the currently largest spectropolarimetric survey of magnetic fields
in several tens of Herbig stars by Alecian et al.\ (2013)
using ESPaDOnS and NARVAL cannot be considered as representative: 
the measurement accuracy in this study is worse than 200\,G for 35\% of the measurements, and 
for 32\% of the measurements it is between 100 and 200\,G.
Clearly, to improve our understanding of the origin of magnetic fields in 
Herbig Ae/Be stars and their interaction with the protoplanetary disk,
it is of utmost importance
to study magnetic fields with high accuracy measurements in a representative 
sample of Herbig Ae/Be stars.

\begin{figure}[t]
\centering
  \includegraphics[width=\textwidth]{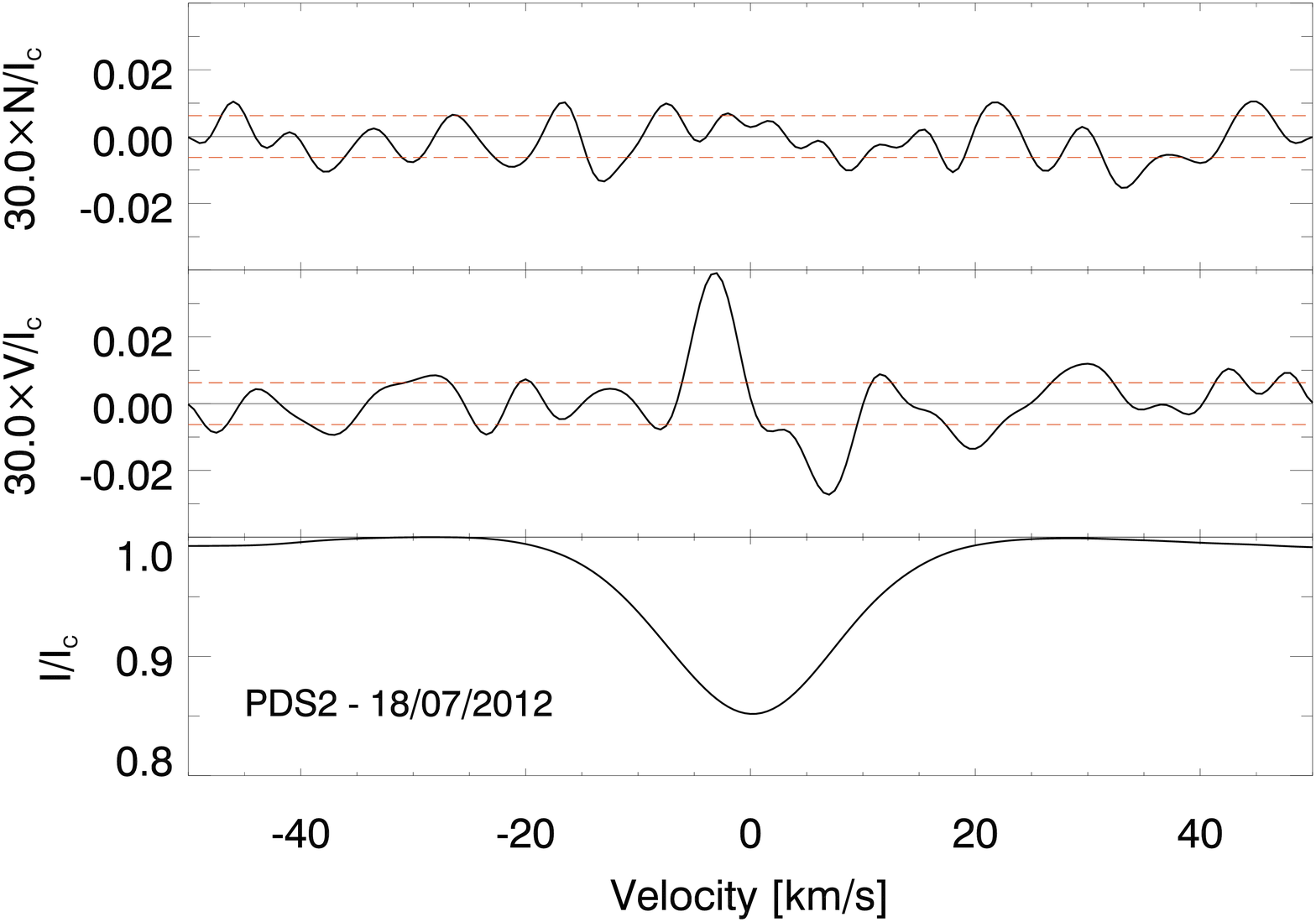}
\caption{
Detection of a mean longitudinal 
magnetic field in the Herbig Ae star PDS\,2. 
The Stokes~$I$ (bottom), Stokes~$V$ (middle), and diagnostic null ($N$; top) SVD profiles were used to determine
$\left<B_z\right>=33\pm5$\,G.
The Stokes~$V$ and $N$ profiles were expanded by a factor of 30.
The red dashed lines indicate the standard deviations for the Stokes~$V$ and $N$ spectra.
This figure corresponds to the lower part of Fig.~2 in
``The prevalence of weak magnetic fields in Herbig~Ae stars: the case of PDS\,2''
by Hubrig et al.\ (2015), MNRAS~449, L118.
}
\label{fig:reduced}
\end{figure}

Zeeman signatures in the spectra of Herbig Ae/Be stars are generally very small, and increasing the 
$S/N$ by increasing the exposure time is frequently limited by the short
rotation period of the star. Therefore, multi-line 
approaches such as LSD and SVD are commonly used to increase the $S/N$.
An example of a very weak Zeeman feature in the HARPS\-pol
spectra of the Herbig Ae star PDS\,2 detected using the SVD method (Hubrig et al.\ 2015) is presented in 
Fig.~\ref{fig:reduced}. 

Spectropolarimetric observations of a sample of 21 Herbig Ae/Be stars observed with FORS\,1
were used to search for a link between the presence of a magnetic field and other stellar
properties by Hubrig et al.\ (2009).
This study did not indicate any correlation of the strength of the longitudinal magnetic 
field with disk orientation, disk geometry, or the presence of a companion. 
No simple dependence on the mass-accretion rate was found, but the range of the observed field values 
qualitatively supported the expectations from magnetospheric accretion models with dipole-like 
field geometries. Both the magnetic field strength and the X-ray emission showed 
a decline with age in the range of $\sim2-14$\,Myr probed by the sample, supporting a dynamo 
mechanism that decays with age. In particular, it was found that magnetic fields are 
strong  in very young Herbig stars and become very weak or completely disappear at the end of 
their PMS life. In line with these results,  Hubrig et al.\ (2000, 2005, 2007) showed that 
magnetic fields in Ap stars with masses less than 3\,$M_{\odot}$ are rarely found 
close to the ZAMS and that kG magnetic fields appear in A stars already evolved from the ZAMS. 
Thus, it is very unlikely that Herbig stars are the progenitors of the magnetic Ap stars.
Importantly, the Herbig stars seem to obey the universal power-law 
relation between magnetic flux and X-ray luminosity established for the Sun and main-sequence 
active dwarf stars (Pevtsov et al.\ 2003). Future work on stellar properties of magnetic Herbig stars 
should involve a larger and more representative sample to determine the existing relations at a higher
confidence level.

As already mentioned in Sect.~\ref{sec:4}, in agreement with the merging scenario for the origin of 
magnetic fields in magnetic stars with radiative
outer zones, the number of close binary systems with Ap components is very low.
Similarly,  studies of Herbig Ae stars
by Wheelwright et al.\ (2010) and Duch\^ene (2015) indicate the lack of close binary systems 
with P$_{\rm orb}<20$\,d. Weak magnetic fields were recently detected in two Herbig Ae systems, in AK\,Sco 
and in the presumed binary HD\,95881.
Using high quality HARPS\-pol observations, J\"arvinen et al.\ (2018)
obtained $\left<B_z\right>=-83\pm31$\,G for the secondary component in the system AK\,Sco and 
$\left<B_z\right>=-93\pm25$\,G for HD\,95881.
It is of interest that for AK\,Sco they observe the magnetic field in the secondary component
in the region of the stellar surface facing permanently the primary component, meaning that the 
magnetic field geometry in the secondary component is likely related to the position of the 
primary component. A similar magnetic field behaviour, where the field orientation is 
linked to the companion, was previously detected in HD\,98088 and HD\,161701, the
two close main-sequence binaries with Ap components mentioned above. 
Obviously, a search for magnetic fields and the determination of their geometries in close binary systems  
is very important as the knowledge of the presence of a magnetic field and of the
alignment of the magnetic axis with respect to the orbital radius vector in Herbig binaries 
may hint at the mechanism of the magnetic field generation.

The weakness of the observed magnetic fields put into question the current understanding of the 
magnetospheric accretion process in intermediate-mass pre-main sequence stars.
Importantly, Cauley \& Johns--Krull (2014) 
studied the He~{\sc i}~$\lambda$10830 morphology in a sample of 
56~Herbig~Ae/Be stars. They suggested that early Herbig~Be stars do not accrete material from their inner
disks in the same manner as T\,Tauri stars, while late Herbig~Be and Herbig~Ae stars show evidence for 
magnetospheric accretion. Furthermore, they proposed more compact magnetospheres in Herbig Ae/Be 
stars compared to T\,Tauri stars.
Ababakr et al.\ (2017) presented  
H$\alpha$ linear spectropolarimetry of a sample of 56~Herbig Ae/Be stars.
A change in linear polarization across this line was detected in 42 (75\%) objects, indicating
that the circumstellar environment around these stars has an asymmetric structure on small spatial scales,
which is typically identified with a disk. A second outcome of their research was 
the confirmation that Herbig Ae stars are similar to T\,Tauri stars in displaying a line polarization 
effect, while depolarization is more common among Herbig Be stars. 
Using near-infrared multi-epoch spectroscopic data obtained with the CRIRES and X-shooter spectrographs 
installed at the VLT,
Sch\"oller et al.\ (2016) examined the magnetospheric accretion in the Herbig~Ae star HD\,101412.
Spectroscopic signatures in He\,{\sc i}~10830 and Pa$\gamma$,
two near-infrared lines that are formed in a Herbig star's accretion region, were presenting temporal modulation.
The authors showed that this modulation is governed by the rotation period of this star and that the observed
spectroscopic variability was explained within the magnetic geometry
established earlier from magnetic field measurements by Hubrig et al.\ (2011).

To summarize, it appears that the as yet small number of magnetic Herbig Ae/Be stars can be due to both
the weakness of the magnetic fields and the large measurement errors. 
According to Alecian (2014), the magnetic properties of Herbig Ae/Be stars
must have been shaped at a very early phase of the stellar evolution.
Using pre-main-sequence evolutionary tracks calculated with the CESAM code  (Morel 1997),
she concluded that even stars above three solar masses will undergo a purely convective phase before 
reaching the birthline.
Therefore, it is reasonable to assume that the weak magnetic fields detected in a number of Herbig Ae/Be stars are
just leftovers of the fields generated by pre-main-sequence dynamos during the convective phase.
If this scenario is valid, we should expect a significantly larger number of Herbig stars possessing weak 
magnetic fields.

\section{Low-mass stars}
\label{sec:6}

Magnetic fields  play a key role in the evolution of solar-like low mass
stars as they determine the angular momentum loss, shape the stellar
wind, and produce high-energy electromagnetic and particle radiation. 
A solar mass star spends approximately 10\,Gyr on the main sequence.
Over the main-sequence life, magnetic activity declines with age and is closely related to a  loss of
angular momentum (e.g.\ Skumanich 1972).
Long-term monitoring of stellar activity using chromospheric emission measured in the cores 
of Ca~{\sc ii} H~\&~K lines revealed that the chromospheric flux increases with the rotation rate 
(Baliunas et al.\ 1995; Hall et al.\ 2007). However, not all active
stars undergo smooth activity cycles and the activity of the most active stars has
the tendency to fluctuate erratically, while regular activity cycles are observed in older stars.
Egeland et al.\ (2017) studied the long-term variability of a sample of five solar analog stars using 
composite chromospheric activity records up to 50~years in length and synoptic visible-band 
photometry about 20 years long. Their sample covered a large range of stellar ages, which they 
used to represent the evolution of activity for solar mass stars. The authors found that young, 
fast rotators have an amplitude of variability many times higher than that of the solar cycle, while old, 
slow rotators have very little variability.

Typical aims in polarimetric studies of low-mass stars are to determine the topology of the surface magnetic 
field and to study the relation between magnetic fields and cool spots.
Most recent studies apply the ZDI technique using Stokes~$IV$ spectropolarimetry and it is
not seldom that no correlation between cool spots and a detected magnetic field is found.
Ros\'en et al.\ (2015) compared ZDI results using Stokes~$IV$ and the full set of 
Stokes~$IVQU$ parameters for the evolved binary RS\,CVn-type variable II\,Peg. 
They found significant differences in the magnetic field solutions and underestimation of 
the total magnetic energy. The retrieval of the meridional field component especially benefited 
from using linear polarization.

Naturally, spectropolarimetric observations of low-mass stars provide excellent constraints 
for theoretical dynamo models. One would expect stars with an internal structure similar to 
that of the Sun to show the same type of dynamo operation. On the other hand, ZDI studies 
frequently find large spots at high latitudes and large regions of a near-surface azimuthal field 
(e.g.\ J\"arvinen et al.\ 2015). 
Petit et al.\ (2008) suggested that the magnetic field is mostly poloidal for low rotation rates and
that more rapid rotators host a large-scale toroidal component in their surface field. 
Further, they inferred that a rotation period lower than about 12\,d is necessary for the toroidal 
magnetic energy to dominate over the poloidal component. Metcalfe et al.\ (2016) proposed that a change in the
character of differential rotation is the mechanism that ultimately disrupts the
large-scale organization of magnetic fields in sun-like stars. The process begins at
a Rossby number (associated with convective motions and presenting the ratio of the rotation
period to the convective turnover time) close to one, 
where the rotation period becomes comparable to the convective turnover
time. The  Vaughan-Preston gap (Vaughan \& Preston 1980) can then be interpreted as a 
signature of rapid magnetic evolution
triggered by a shift in the character of differential rotation, which is an emergent property of 
turbulent convection in the presence of Coriolis forces.

The red dwarfs are the coolest objects which include late K and M dwarfs. 
The majority of red dwarfs exhibit evidence of strong magnetic activity,
which is expressed in strong optical flares, enhanced UV, X-ray, and radio emission. 
The process of generating strong magnetic fields
in fully convective objects and those with very deep convection zones is not yet understood.
First direct magnetic field measurements in these dwarfs
were achieved using the Zeeman split Fe~{\sc i} line at 8468.4\,\AA{} (Johns-Krull \& Valenti 1996).
More recent spectropolarimetric observations demonstrated that early type 
M dwarfs have magnetic fields with a strong
toroidal  component,  reminiscent  of those  of active K and G dwarfs,  whereas  the  
lowest-mass M dwarfs exhibit magnetic fields that are highly
organized and strongly poloidal (e.g.\ Morin et al.\ 2008a) and some of them show indications of polar
starspot features (Morin et al.\ 2008b).

Starspots are the primary evidence of magnetic activity harboring the strongest field and the coldest plasma.
Spectropolarimetric surveys of red dwarfs revealed the presence of numerous atomic and molecular signatures
formed in starspots and the chromosphere. First detections of polarization in TiO, CaH, and FeH  transitions in 
starspots
on M and K dwarfs were presented by Berdyugina et al.\ (2006, 2008) using ESPaDOnS.
It was the first time that polarimetry of cool atmospheres with the help of molecular 
physics was used to obtain new insights into the magnetic structures of cool atmospheres.
For the red dwarf AU\,Mic, the authors identified four spots,
one of positive and three of negative polarity. These spots are 500--700\,K cooler than the photosphere
and harbour a maximum magnetic field of 5.3\,kG.

Brown dwarfs, often referred to as ``failed stars'' because they do not sustain hydrogen fusion in their core,
show magnetic activity similar to that of low-mass stars and produce huge flares (Paudel et al.\ 2018).
The first direct detection of a strong, 5\,kG magnetic field on the surface of an active brown dwarf
was presented in the work of Berdyugina et al.\ (2017). The surface magnetic
field of the brown dwarf with a mass of about 55\,$M_{\rm J}$ and at an age of 22\,Myr 
was detected in circularly polarized signatures in the 819\,nm sodium line. 
According to the authors, the young age of the magnetic brown dwarf implies that it may still maintain a disk,
which may facilitate bursts via magnetospheric accretion, like in higher-mass T\,Tauri stars.

The study of magnetic fields of low-mass stars
became particularly important after the discovery of exoplanets around such stars.
Multiwavelength observations with different observing strategies
concluded that a giant exoplanet in a short-period
orbit can induce activity in the photosphere and upper atmosphere of its parent star.
This makes the host star's magnetic activity a probe of the planet's magnetic field.
The large scale magnetic field of the planet-host star  HD\,189733 with a mainly toroidal surface magnetic field
with  a  strength  of  20--40\,G has been observed over multiple
years. The ZDI analysis of these observations revealed a structural evolution of the field
between observations (e.g.\ Fares et al.\ 2013).

\section{Post-main-sequence stars}
\label{sec:7}

The evolved binary components in RS\,CVn-type variables show remarkable activity, displaying strong
chromospheric plages, coronal X-ray and microwave emission and flares detected in the optical,
UV, radio, and X-rays.  For these systems, magnetic activity cycles have been suggested to be the cause of
the detected period modulations. One of the RS\,CVn systems exhibiting substantial period variations
is HR\,1099 with a high level of magnetic activity in all available activity indicators.
Studies using photometric observations reported different  magnetic activity periods,
from 14.1\,yr (Muneer et al.\ 2010) to 19.5\,yr (Lanza et al.\ 2006).
The Applegate mechanism proposed by Applegate (1992)
has recently been discussed by V\"olschow et al.\ (2018) as a potential intrinsic
mechanism to explain transit timing variations in various kinds of close binary systems. 
In the framework of the Applegate model,
period modulations can be explained by gravitational coupling of the orbit to variations in 
the shape of the magnetically active star in the system. The variable deformation of the active star 
is produced by variations in the distribution of angular momentum as the star goes through its 
activity cycle. The torque needed to redistribute the angular momentum can be exerted by a mean 
subsurface magnetic field of several kilogauss. 

In former studies,
an exciting topic of the magnetic activity of RS\,CVn-type variables and 
the FK\,Comae stellar group was the flip-flop phenomenon. 
The FK\,Comae stellar group consists of single G--K giants showing 
high activity and rapid rotation.
In many such active stars, the spots were found to concentrate on two 
permanent active longitudes, 
which were half a period apart (see e.g.\ Berdyugina et al.\ 1998). In some of these stars, 
the dominant part of the spot activity concentrated on one of the active longitudes, and abruptly 
switched the longitude every few years. This flip-flop phenomenon was discovered in 
FK\,Comae (Jetsu et al.\ 1993). Light curves and Doppler imaging showed evidence of 
differential rotation of modest strength. However, in his most recent work, Jetsu (2018) 
showed that the period search methods used in previous studies do not necessarily produce correct rotation 
period estimates, if there are multiple interfering periodicities. 
Based on his new two-dimensional period finding method, Jetsu argued that the observed deceptive light 
curves were the interference of two real constant period light 
curves of long-lived starspots. The slow motion of these long-lived starspots with respect to each 
other caused the observed light curve changes. 

Hubrig et al.\ (1994) carried out magnetic field measurements for thirteen bright G--K giants of different X-ray 
luminosity and concluded that magnetic fields in such objects are indeed observable using spectropolarimetry.
Later on, direct detections of magnetic fields in the photospheres of cool, evolved 
giants advancing on the red giant branch and the asymptotic giant branch  were presented by Konstantinova-Antova et al.\ (2013) and Auri\`ere 
et al.\ (2015). In the sample of 48 red giants studied by Auri\`ere et al.,
magnetic fields were detected via Zeeman signatures in 29 giants.
The majority of the magnetic giants are either in the first dredge up
phase or at the beginning of core He burning, i.e.\ in phases when the convective turnover time is at a maximum.
Four giants for which the magnetic field was measured to be outstandingly strong, up to 
100\,G (with respect to that
expected from the relation between rotation period and magnetic field or from their evolutionary status), were interpreted 
as being probable descendants of magnetic Ap stars.

\begin{figure*}[t]
\centering
 \includegraphics[width=0.8\textwidth]{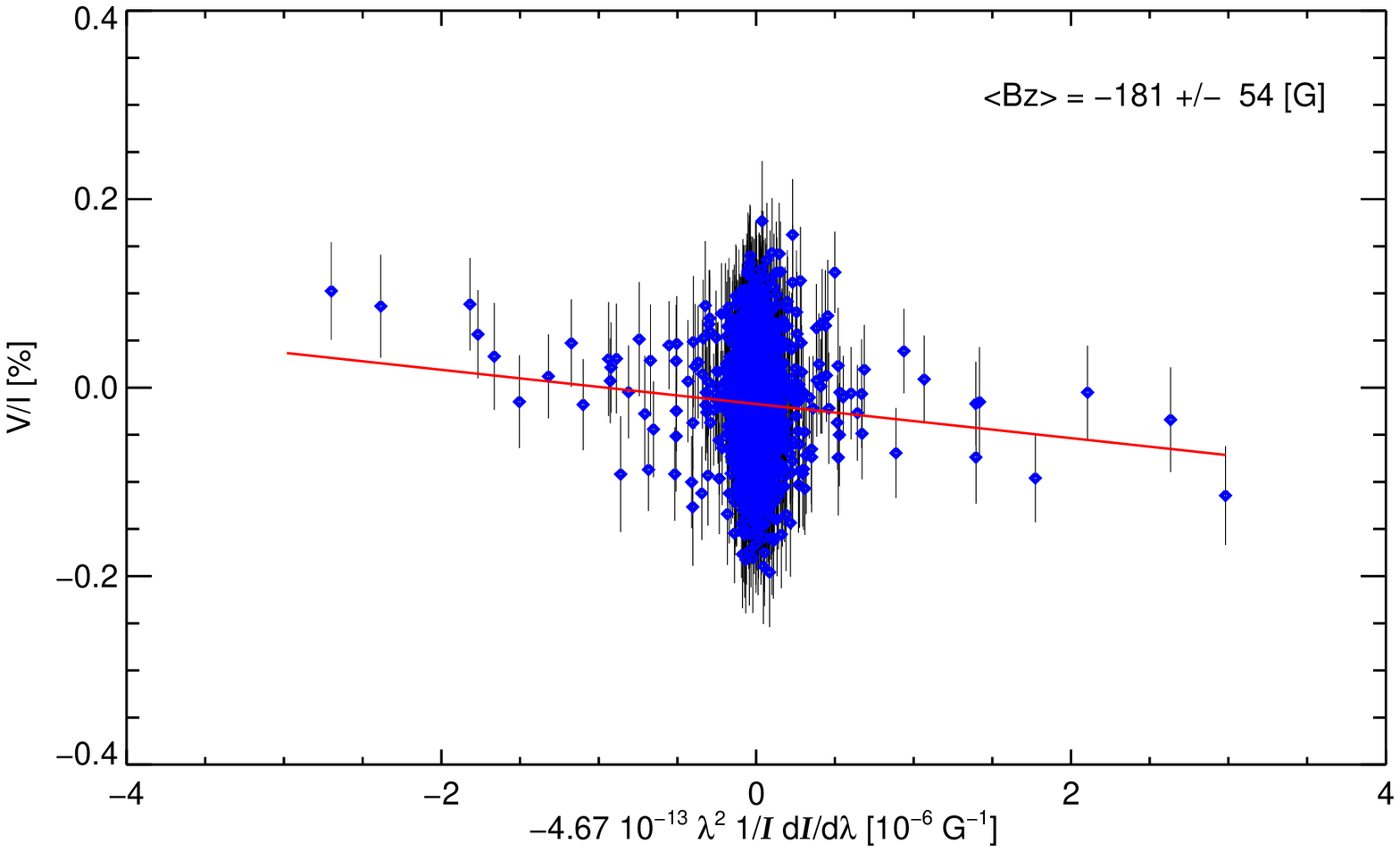}
 \includegraphics[width=0.8\textwidth]{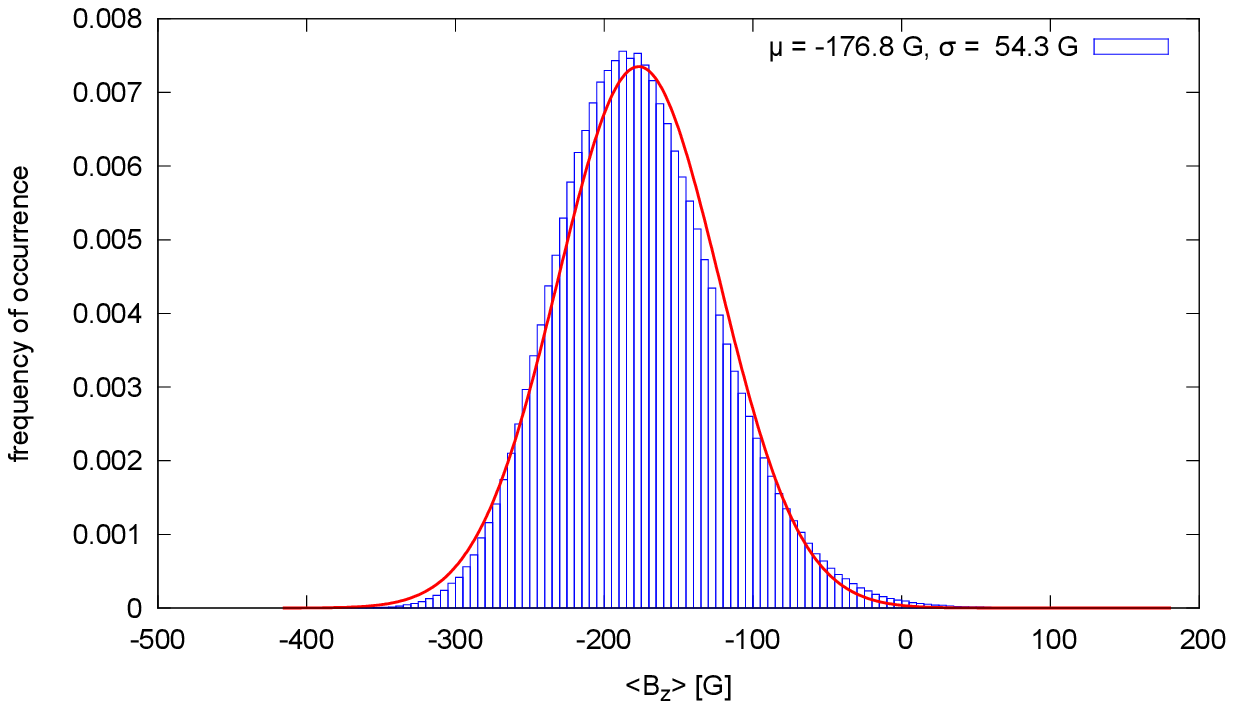}
\caption{
IC\,418: Regression detection of a 
$\left<B_{\rm z}\right>=-181\pm54$\,G mean longitudinal magnetic field, 
using all lines (top). The corresponding 
$\left<B_{\rm z}\right>$ distribution (bottom) 
deviates slightly from a Gaussian (red curve) and indicates 
$\left<B_{\rm z}\right>=-177\pm54$\,G. Figure from Steffen et al.\ (2014, A\&A, 570, A88);
reproduced with permission, \copyright{} ESO.
}
\label{fig:neb}
\end{figure*}

One of the major open questions regarding the formation of planetary nebulae (PNe) concerns the 
mechanism that is responsible for their non-spherical, often axisymmetric shaping (e.g.\ 
Balick \& Frank 2002). Both, central star binarity and stellar
rotation in combination with magnetic fields, are among the favourite explanations.
Basically, the origin of PNe is understood to be a consequence  of  the  interaction  of  the  
hot  central  star  with  its  circumstellar environment through photoionization and wind-wind
collision. 
In principle, the role of magnetic fields in shaping PNe  may be verified or 
disproved by empirical evidence, as already suggested by Jordan et 
al.\ (2005).
Using FORS\,1 in spectropolarimetric mode, they reported the detection of 
magnetic fields of the order of kG in the central stars of 
the PNe NGC\,1360 and LSS\,1362. 
A reanalysis of their data, however, did not provide any significant evidence 
for longitudinal magnetic fields that are stronger than a few 
hundred Gauss in these stars (Jordan et al.\ 2012).
Their field measurements had typical error bars of 150 to 300\,G.
Similar results were achieved in the work of Leone et 
al.\ (2011), who concluded that the mean longitudinal magnetic 
fields in NGC\,1360 and LSS\,1362 are much weaker,
less than 600\,G, or that the magnetic field has a complex structure.
The most recent search for magnetic fields in central stars of planetary 
nebulae by Leone et al.\ (2014), based on spectropolarimetric 
observations of 19 central stars with the Intermediate-dispersion Spectrograph and Imaging System (ISIS)
on the William Herschel Telescope (WHT) and VLT/FORS\,2, was partly
affected by large measurement uncertainties and reported
no positive detection either.

Using low-resolution polarimetric spectra obtained with FORS\,2, Steffen et al.\ (2014)
carried out a search for magnetic fields in a
sample of 12 central stars covering the whole range of morphologies from 
round to elliptical/axisymmetric, and bipolar PNe, and including both 
chemically normal and Wolf-Rayet type (hydrogen-poor) central stars. The 
sample included two round nebulae (NGC\,246, Hen\,2-108), five elliptical 
nebulae (IC\,418, NGC\,1514, NGC\,2392, NGC\,3132, Hen\,2-131), and three
bipolar nebulae (NGC\,2346, Hen\,2-36, Hen\,2-113). Two targets have 
WR-type central stars (NGC\,246, Hen\,2-113). In addition, the authors included 
the two (elliptical) targets of Jordan et al.\ (2005), 
NGC\,1360 and LSS\,1362, for which they originally claimed the detection of 
kG magnetic fields. Six of the 12 central stars were known binaries.

Formal 3$\sigma$ detections were achieved for IC\,418 (${\left<B_{\rm z}\right>=-181\pm54}$\,G), for 
the WR-type central star Hen\,2-113 (${\left<B_{\rm z}\right>=-58\pm18}$\,G), and 
the weak emission line star Hen\,2-131 (${\left<B_{\rm z}\right>=-120\pm32}$\,G).
The mean longitudinal magnetic field $\left<B_{\rm z}\right>$ of the central stars was derived by 
linear regression employing two different methods:
a) using  a weighted linear regression line through the measured data points
and b) generating $M=10^6$ statistical variations of the 
original dataset by the bootstrapping technique and analyzing the resulting 
distribution $P(\left<B_{\rm z}\right>)$ of the $M$ regression results. Mean 
and standard deviation of this distribution are identified with the most 
likely mean longitudinal magnetic field and its 1$\sigma$ error, 
respectively. The main advantage of this method is that it provides an 
independent error estimate.
The results of both methods for the central star of the young, elliptical planetary 
nebula IC\,418 are presented in Fig.~\ref{fig:neb}. A mean longitudinal magnetic field of 
negative polarity was detected at a 3$\sigma$ significance level when using the entire 
spectrum: ${\left<B_{\rm z}\right>=-181\pm54}$\,G using the first method and
${\left<B_{\rm z}\right>=-177\pm54}$\,G for the second method. 
Most stars in the sample of Steffen et al.\ (2014) were not studied at the achieved
accuracy before, permitting to put 
constraints on the strength of the magnetic fields in the central stars
of planetary nebulae.

\section{Degenerate stellar remnants}
\label{sec:8}

White dwarfs represent the final stage of stellar evolution for the majority of
stars  with masses less than about 8\,$M_{\odot}$.
The majority of white dwarfs have hydrogen-rich atmospheres (DA). Other types of 
white dwarfs include stars with atmospheres dominated by helium (DB) and those without
a visual helium spectrum (DC). 25\% to 30\% of white dwarfs show
traces of heavy elements, i.e.\ elements heavier than helium (Zuckerman et al.\ 2003, 2010), probably due to 
accreted planetary or asteroidal debris (Jura 2008). 

A significant fraction of white dwarfs possess a magnetic field ranging from
a few kG up to several hundred MG (e.g.\ Ferrario et al.\ 2015). 
According to Kepler et al.\ (2013), the incidence of magnetism within the white dwarf
population remains uncertain with an incidence of magnetism of about 5\% in magnitude limited surveys,
or between 10--20\% in volume limited surveys (e.g.\ Kawka et al.\ 2007).
Kawka \& Vennes (2014) showed that  cool, polluted white dwarfs have a higher incidence of 
magnetism, up to 40\%, than in the general population of white dwarfs. Dufour et al.\ (2013) 
reported on a magnetic field incidence of about 70\% 
in the rare class of warm and hot DQ white dwarfs, which present rare carbon-dominated objects 
known as hot DQ white dwarfs.

To study magnetic fields in white dwarfs, again circular polarization observations are involved. Similar 
to Ap and Bp stars, the geometry of the magnetic field in white dwarfs is assumed to be a
centred or offset dipole.
However, the magnetic field structure in white dwarfs studied over their rotational period 
shows a large diversity, from a simple offset dipole, over structures with spots,
to multipoles (e.g.\ Kawka et al.\ 2018). For non-variable magnetic white dwarfs 
it is assumed that they are either rotating with a very long period
or that their magnetic fields are nearly aligned with the rotation axis. 
A recent statistical analysis of the linear polarization properties between magnetic and non-magnetic
white dwarfs showed no difference in the polarization degrees ({{\.Z}ejmo} et al.\ 2017; {S{\l}owikowska} et al.\ 2018).
Such an analysis allowed the authors to select a set of good candidates of faint linear polarimetric 
standard stars that can be used as stable polarimetric calibration sources.

The mechanism responsible for the presence of magnetic fields in these stars
remains unclear, as white dwarfs do not contain any significant convective zones.
In the framework of the fossil field theory for the origin of magnetic fields,
and assuming magnetic flux conservation, Ap and Bp stars are unlikely
progenitors of white dwarfs with weak magnetic fields. 
Tout et al.\ (2008) proposed a binary origin, where the magnetic field is formed
via a dynamo created during a common envelope phase.
Nordhaus et al.\ (2011) proposed that
during a CE phase a low-mass star will be tidally disrupted by its proto-white
dwarf companion, forming an accretion disk and generating a dynamo in
the disk, which is then transferred to the degenerate core via accretion. 
Garc\'ia-Berro (2012) suggested that magnetic
fields can also be produced by the merger of two white dwarfs. 
To explain the sharp increase in the incidence of magnetism
in cool polluted white dwarfs,
Kawka \& Vennes (2014) argued that the phenomenon coincides with the likely presence of a 
perennially crowded circumstellar environment, which may have been caused by the same event that 
generated the magnetic field.
As various classes of white dwarfs show a different incidence of magnetic fields, it is plausible
that magnetic white dwarfs are created by several discernible processes.

For a discussion of the magnetic properties of neutron stars see Chapter~6. 

%
%
%

\end{document}